# *PLS-SEM-power*: A Shiny App and R package for Computing Required Sample Size and Minimum Detectable Effect Size in PLS-SEMs

Alessandro Ansani[1]✉

[1] Centre of Excellence in Music, Mind, Body and Brain (CoE MMBB) – Department of Music, Art and Culture Studies, University of Jyväskylä (Jyväskylä, Finland)

✉ *Correspondence*:

Alessandro Ansani, PhD
Department of Music, Art and Culture Studies | University of Jyväskylä
Musica Building M - Seminaarinkatu 15, 40014 – Jyväskylä, Finland

## Abstract

Despite its evanescent nature, statistical power is crucial for planning Partial Least Squares Structural Equation Modelling (PLS-SEM) studies. This brief paper introduces *PLS-SEM-power*, a Shiny Application and R package that implements the inverse square root method by Kock and Hadaya (2018) to calculate both the minimum required sample size (*a priori analysis*) and the Minimum Detectable Effect Size (MDES, *sensitivity analysis*), given a chosen significance level ($\alpha$ level) and desired statistical power ($1 - \beta$), set to 80% by default. The software supports both one-tailed and two-tailed calculations, with one-tailed calculations as the default following the formulation of Kock and Hadaya (2018), and provides an intuitive user interface, facilitating reproducible and easily accessible analyses in diverse research contexts.

# Introduction

The interest in Partial Least Squares Structural Equation Models (PLS-SEMs) from the research community has been growing exponentially over the last ten years, according to Google Trends and Scopus [Fig. 1].

**Figure 1**. Growth of Google searches and Scopus documents over time

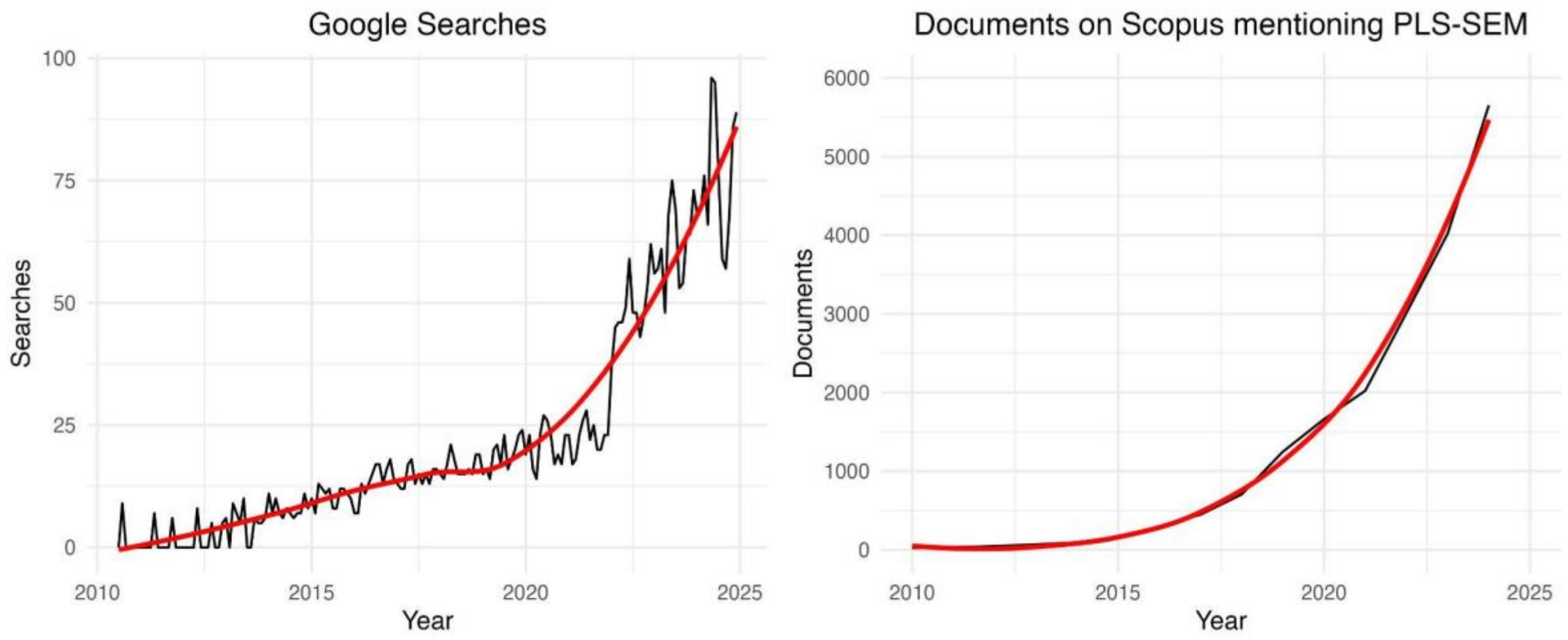

Several factors explain such growth. These include, for instance, methodological and technological advantages, besides a fundamental philosophical-theoretical difference, making the methodology particularly well-suited to a wide range of study situations in social and management science. Herman Wold's creation of PLS-SEM in the 1970s (Hair et al., 2018) provided researchers with an alternative to Covariance-Based Structural Equation Modelling (CB-SEM). However, broad acceptance began in the 2000s, when accessible software became available. Tools such as *PLS-Graph* (Chin, 2001) and later *SmartPLS* (Ringle et al., 2024) democratised the technique, allowing researchers without considerable statistical experience to use these methods quite efficiently. Also known as “soft” design models compared to CB-SEMs, often labelled as “hard modelling”, PLS-SEMs are valued for their flexibility. Unlike CB-SEM, which requires strong assumptions (multivariate normality, large samples, well-specified patterns), PLS-SEMs, in fact, manage to work with non-normal data, smaller samples (Rigdon, 2016), complex patterns, and both reflexive and formative measurements (Hair et al., 2011). This has made PLS-SEMs particularly useful in disciplines such as marketing (e.g., Fornell & Bookstein, 1982; Henseler et al., 2009), medicine, and psychology (Willaby et al., 2015), where data do not often meet classical assumptions. Perhaps most significantly, PLS-SEM prioritises prediction over explanation. Rather than seeking to reproduce observed covariances, it maximises the variance explained ($R^2$) in dependent constructs. This orientation makes it particularly valuable for forecasting and theory building in emerging research areas (Sarstedt et al., 2014). In contrast, CB-SEM focuses on covariance explanation and cannot generate unique latent variable scores, constraining its predictive applications (Dijkstra, 2014).

## The issue of power in PLS-SEM

In the design of quantitative studies, researchers must always consider the sample size needed to achieve sufficient statistical power for the hypothesised effect size. In this respect, one of the most convenient features of PLS-SEM is that the model structural complexity does not have a critical impact on the determination of the minimum sample size required (Hair et al., 2021, Chapter 1). Moreover, PLS-SEM has been argued to offer greater statistical power than CB-SEM under conditions in which the assumptions of the latter are violated, such as small sample sizes, nonnormal data, and high model complexity (Hair et al., 2011, p. 143). However, this potential advantage is not unequivocally supported by empirical evidence. Based on Monte Carlo simulations, Goodhue and colleagues (2012) found no consistent advantage of PLS over CB-SEM in terms of statistical power under these conditions, with all approaches exhibiting reduced power as sample sizes decreased. These findings suggest that PLS-SEM should not be regarded as a remedy for inadequate sample sizes and that formal power analysis remains essential, regardless of the SEM approach adopted.

As described by Hair and colleagues (2021), these convenient features led to the misconception that sample size computation has no relevance in PLS-SEM. This misconception has been fueled by the very common '10-times rule' (Barclay et al., 1995), i.e., a heuristic often used to determine the minimum sample size required for a study (Hair et al., 2011). According to this rule, the sample size "should be equal to 10 times the number of independent variables in the most complex regression in the PLS path model (i.e., considering both measurement and structural models). This rule of thumb is equivalent to saying the minimum sample size should be 10 times the maximum number of arrowheads pointing at a latent variable anywhere in the PLS path mode." (Hair et al., 2021, p. 16). This approach has the undeniable merit of being simple and easy to apply, which has contributed to its widespread use. There are nevertheless numerous problems (Goodhue et al., 2012). First, the rule is overly simplistic; it does not account for other critical factors that influence statistical power, such as the effect size, significance level ($\alpha$), and the desired statistical power ($1-\beta$). The fact that this strategy is not sensitive to the effect size, in particular, is problematic. The basic underlying principle behind every power analysis, regardless of the specific technique at hand, is precisely that the sample size must vary based on the magnitude of the hypothesised effect. As a result, the 10-times rule may underestimate the required sample size, thus resulting in underpowered studies. As a second point, the 10-times rule lacks a solid statistical foundation. It relies on informal and universal guidelines rather than rigorous mathematical derivations, thus limiting its reliability and applicability across a broad range of research scenarios.

The Minimum R-squared method (Hair et al., 2014) represents an alternative approach to sample size calculation. Using the smallest R-squared value anticipated in the model, this method computes the required sample size in PLS-SEM. By calculating the minimum R-squared, the corresponding sample size needed to achieve statistical power at a given significance level ($\alpha$) is determined using either a lookup table or a formula. Despite being simpler, accounting for the strength of relationships in the model, and having stronger statistical foundations, this approach has a number of theoretical and practical limitations. On the theoretical side, the $R^2$ is a measure of the model's explanatory power (Shmueli & Koppius, 2011), more correctly referred to as in-sample predictive power (Rigdon, 2012, but see also Hair et al., 2021, chapter 6). However, the predictive endeavour is hardly the main aim in the social sciences (Ozili, 2023), wherein the ultimate goal is unbiased causal inference. Moreover, despite the existence of some commonly used thresholds for acceptable $R^2$ values (Hair et al., 2011), they are not entirely agreed upon in the literature, and it has been shown convincingly that models with $R^2$ values as low as 0.10 can be entirely satisfactory in certain scientific contexts (Raithel et al., 2012). Furthermore, the idea that high predictive power directly translates into explanatory power and more legitimate causality claims is a false belief (McElreath, 2020, p. 225; Hill et al., 2018). On the practical side, the Minimum R-squared method assumes that researchers have a rough idea of, or can even quite accurately predict, the minimum $R^2$ prior to data collection. This is oftentimes far from feasible, especially in the social sciences realm, and even more so when assessing the relationships among more novel or rarely analysed psychological constructs.

For these and other reasons, Kock and Hadaya (2018) introduced the *inverse square root method*, a versatile tool applicable in various research contexts. This novel approach determines the minimum sample size by exploiting the inverse square root relationship between the sample size and the standard error of the path coefficient estimates. The method is based on a simple formula that integrates a critical value constant ($p_\alpha$), derived from the chosen significance level (α) and power (1−β). Thanks to its computational efficiency, this method provides clear and immediate guidance for estimating the sample size needed to ensure reliable results in PLS-SEM. Furthermore, its only requirement is for the researcher to have an idea of the effect size(s) at hand, and the desired α level.

## The Formula and Problem Statement

The core formula to find the minimum required sample size ($N$) for a given effect size (Equation 5 in Kock & Hadaya, 2018) is:

$$N = \left(\frac{p_\alpha}{p_{min}}\right)^2$$

Where the numerator, $p_\alpha$, is a constant which depends on the α and (1−β) level, namely:

$$p_\alpha = z(1-\alpha) + z(1-\beta)$$

For example, with 1−β = 0.80 and α = 0.01, $p_\alpha$ = 3.168; with α = 0.05, $p_\alpha$ = 2.486; with α = 0.10, $p_\alpha$ = 2.123).

The denominator, $p_{\min}$, is the value of the focal path coefficient with the minimum size in the path model that is expected to be statistically significant.

When $N$ is given, we can instead solve the equation for $p_{\min}$, the minimum magnitude of the path coefficient which is expected to be statistically significant. Rearranging the formula, $p_{\min}$ is expressed as:

$$p_{min} = \frac{p_\alpha}{\sqrt{N}}$$

where $N$ is the known sample size and $p_\alpha$ remains the constant corresponding to the chosen significance level (α).

This simple rearranged formula provides a practical approach to determine the minimum effect size that can be detected at a specified level of statistical power ($1 - \beta$), significance level ($\alpha$), and sample size. In the literature, this is often referred to as Minimum Detectable Effect Size (MDES: Bloom, 1995; Dong & Maynard, 2013).

It is worth noting that the values of $p_\alpha$ reported above correspond to the one-tailed formulation used by default in *PLS-SEM-power*, following Kock and Hadaya (2018)[1]. Two-tailed calculations are also available, following the approach discussed by Kock (2015), to which readers are referred for a broader discussion of one- versus two-tailed tests in PLS-SEMs.

### Existing Software Solutions

Existing software provides several options for sample size and power analysis in PLS-SEM. *WarpPLS* (Kock, 2025) implements both the inverse square root and gamma-exponential methods, while *SmartPLS* (Ringle et al., 2024) provides a post hoc minimum sample size assessment based on the inverse square root method. The R package *semPower* (Moshagen & Bader, 2023) offers a broader framework for power analysis in structural equation modelling, using model-based approaches rather than the inverse square root method. In this context, *PLS-SEM-power* provides a lightweight, open-source implementation of the inverse square root method, allowing researchers to conduct both a priori and sensitivity analyses without requiring a dedicated PLS-SEM software environment. Its accompanying Shiny application further extends accessibility to users without programming experience.

## PLS-SEM-power: Description of the Shiny App

To make these calculations more accessible and reproducible, we developed PLS-SEM-power. The application can be accessed freely at *https://aleansani.shinyapps.io/pls-sem-power*. Aside from a standard web browser, no additional software installations are required.

The application provides two primary modes of use:

1. ***A priori* (given MDES, compute sample size):** The researcher selects an α level (from .01, .05, or .10) and specifies the target MDES, selects either a one-tailed or two-tailed calculation, and sets the desired statistical power (0.80 by default). The app computes and displays the minimum required sample size.
2. ***Sensitivity* (given sample size, compute MDES):** The researcher selects an α level (from .01, .05, or .10) and specifies the sample size, selects either a one-tailed or two-tailed calculation, and sets the desired statistical power (0.80 by default). The app calculates and displays the Minimum Detectable Effect Size (MDES).

[1] For two-tailed tests, the constants are calculated as $p_\alpha = z(1 - \frac{\alpha}{2}) + z(0.80)$, assuming 80% statistical power. The corresponding values are 3.418, 2.801, and 2.525 for $\alpha$ = .01, .05, and .10, respectively.

### Interface and Usage

- **α selection**: A set of radio buttons allows the user to select the significance level from .01, .05, or .10.
- **Tail selection**: Two radio buttons allow the user to select a one-tailed or two-tailed calculation. One-tailed calculations are selected by default, following Kock and Hadaya (2018).
- **Statistical Power**: The user enters the desired value of statistical power (1 – β).
- **Data Input**: Depending on the selected analysis (*A priori* or *Sensitivity*), the user enters the desired MDES or the sample size.
- **Results**: The main panel displays the calculated sample size or MDES using the formulas above and indicates whether a one-tailed or two-tailed calculation was used.
- **Graphs**: Two interactive plots (*A priori* and *Sensitivity*) illustrate the relationship between MDES and sample size. Reference lines (vertical and horizontal) indicate the user-defined point of interest. The graphs use Okabe and Ito's (2008) colour-blind-friendly colour palette. Users can choose between two plot themes: *abyss* (default), a dark theme, and *min* or *pr*, a publication-ready B&W theme.

## PLS-SEM-power: Description of the R package

To facilitate R users who wish to implement *PLS-SEM-power* locally, we developed an R package parallel to the Shiny App, with identical functions.

### Installation

The package requires R (≥ 4.1.0). The installation is straightforward and runnable through the following code:

```
 # install.packages("devtools") # if needed
library(devtools)

 # install plssempower
devtools::install_github("AleAnsani/plssempower")
```

### Functions

Similarly to the Shiny App, *plssempower* provides two simple functions:

- `pls_sem_power()` – Computes either the minimum sample size needed for a given Minimum Detectable Effect Size (MDES), or the MDES for a given sample size, based on the formula proposed by Kock and Hadaya (2018).
- `pls_sem_power_graph()` – Generates ggplot2 graphs showing the relationship between MDES and sample size, including reference lines for the user's input.

For both functions, the user can choose between a one- or two-tailed calculation via the `tail` (or `tails`) argument. Furthermore, they can enter the desired level of statistical power (0.80 by default).

### Usage

```
  # Compute the required sample size for a given MDES
pls_sem_power(method = "a priori", MDES = 0.5, alpha = 0.05, tail = 1,
power = .80)

  # Compute MDES for a given sample size
pls_sem_power(method = "sensitivity", N = 68, alpha = 0.05, tail = 1,
power = .80)

  # Generate graph (a priori)
pls_sem_power_graph(method = "a priori", MDES = 0.5, alpha = 0.05,
tail = 1, power = .80)

  # Generate graph (sensitivity)
pls_sem_power_graph(method = "sensitivity", N = 68, alpha = 0.05, tail
= 1, power = .80)
```

### Examples

*A priori*

Assume the researcher wishes to investigate a specific path within a PLS-SEM for which the expected effect size is approximately 0.5. In such a case, an *a priori* power analysis is required to determine the minimum sample size needed to detect the hypothesised effect with 80% statistical power. This analysis can be carried out simply by running the following command:

```
pls_sem_power_graph(method = "a priori", MDES = 0.5, alpha = 0.05, tail
= 1, power = .80)
```

The resulting plot, as shown in Figure 2, shows that, with a statistical power of 80%, a minimum of 25 participants is needed to detect the designated effect size using a one-tailed test.

**Figure 2**. Example of an *a priori* power analysis graph with PLS-SEM-power

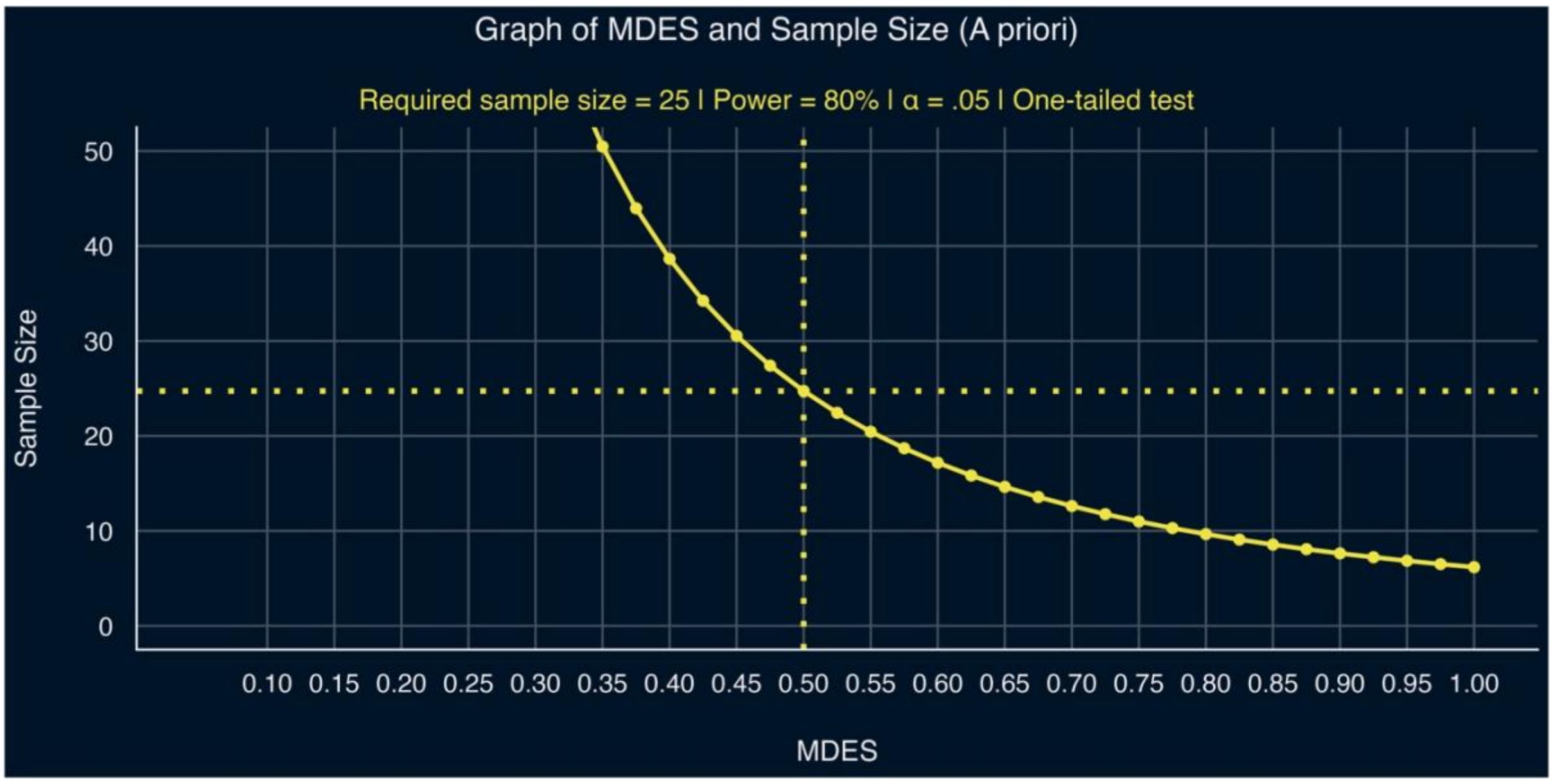

For users who do not require graphical output and are solely interested in obtaining the numerical result, the sample size calculation can also be performed by executing the following command:

```
pls_sem_power(method = "a priori", MDES = 0.5, alpha = 0.05, tail = 1,
power = .80)
```

The function returns the following message:

```
To detect an effect of 0.5 with 80% power at alpha = 0.05 using a one-
tailed test, you need at least 25 observations.
```

In case the researcher does not have a specific hypothesis about the direction (positive or negative) of the effect (see Kock, 2015), it is possible to perform a two-tailed calculation by simply setting:

```
pls_sem_power(method = "a priori", MDES = 0.5, alpha = 0.05, tail = 2,
power = .80)
```

In this case, the function returns the following message:

```
To detect an effect of 0.5 with 80 % power at alpha = 0.05 using a
two-tailed test, you need at least 32 observations.
```

Thus, in this specific case, switching from a one-tailed to a two-tailed calculation increases the required sample size by seven observations, corresponding to a 28% increase relative to the one-tailed requirement.

The research might also want to increase the desired level of statistical power, for instance:

```
pls_sem_power(method = "a priori", MDES = 0.5, alpha = 0.05, tail = 2,
power = .90)
```

In that case, the required sample size increases:

```
To detect an effect of 0.5 with 90 % power at alpha = 0.05 using a
two-tailed test, you need at least 43 observations.
```

*Sensitivity*

For the *sensitivity* scenario, suppose the researcher has already collected the data prior to conducting any power analysis. In this example, the final sample consists of 68 participants. The researcher may now wish to determine the smallest effect size that can be reliably detected with a statistical power of 80% using a one-tailed test. Although this approach is less optimal than conducting an *a priori* power analysis, carrying out an *a priori* power analysis is still of help to make the research process more transparent.

Given the available sample size, a sensitivity power analysis is suitable in these situations to estimate the MDES. The following command can be used to achieve this:

```
pls_sem_power_graph(method = "sensitivity", N = 68, alpha = 0.05, tail
= 1, power = .80)
```

The function works in the same way as the *a priori* version; the only necessary adjustments involve indicating the method as "sensitivity" and supplying the sample size (*N*) rather than the expected effect size (MDES).

As shown in Figure 3, the resulting graph indicates that, with such a sample size, effects greater than 0.302 can be detected with 80% power using the default one-tailed calculation.

**Figure 3**. Example of a *sensitivity* power analysis graph with PLS-SEM-power

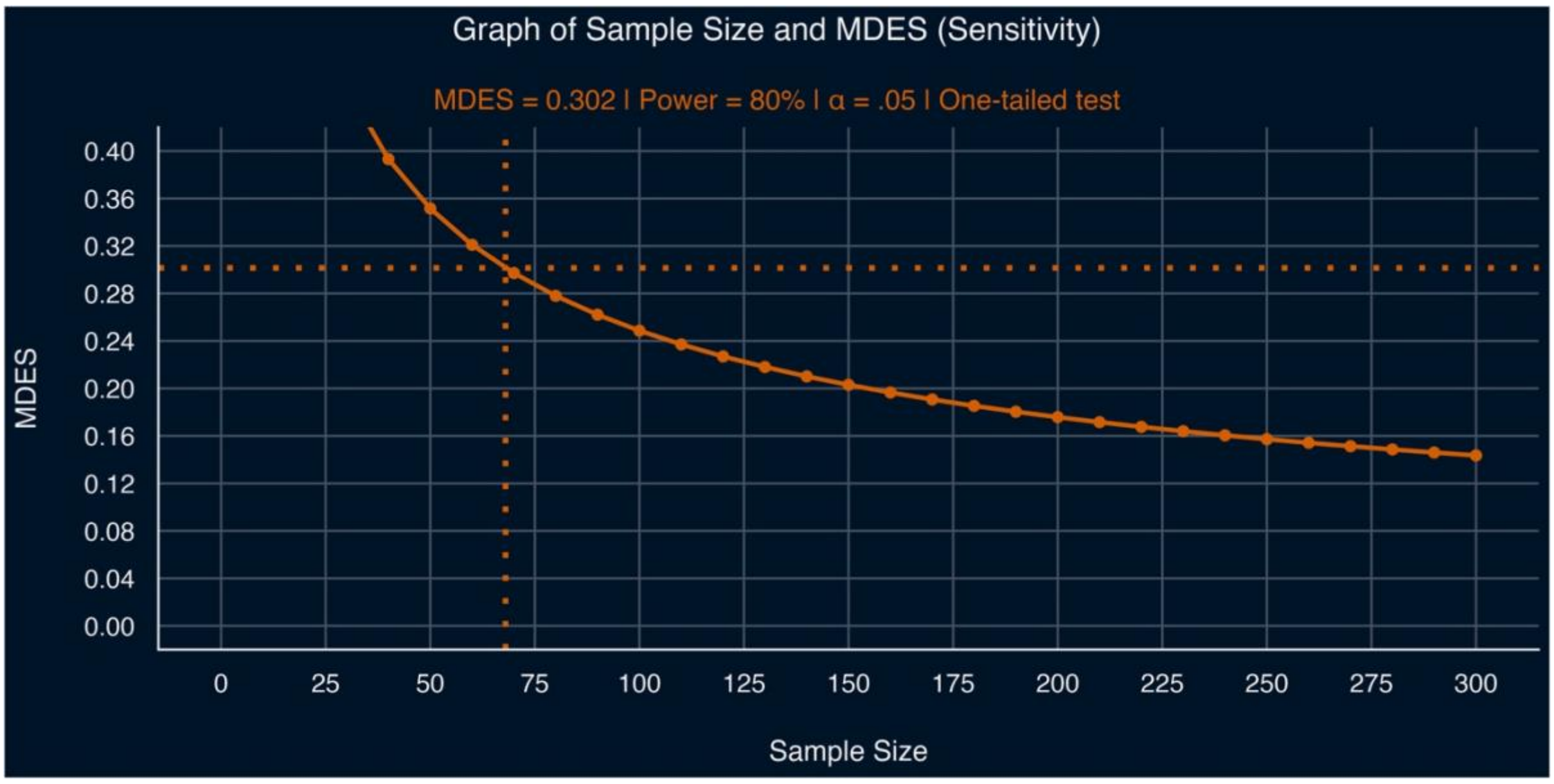

For users who prefer a textual output and do not require graphical visualisation, the same analysis can be performed with:

```
pls_sem_power(method = "sensitivity", N = 68, alpha = 0.05, tail = 1,
power = .80)
```

The function returns the following message:

```
With N = 68 and alpha = 0.05 using a one-tailed test, you can detect
effects as small as 0.302 with 80% power.
```

Finally, users who wish to generate a black-and-white, publication-ready graph can use the "pr" theme by specifying `theme = "pr"` in the function call. For example:

```
pls_sem_power_graph(method = "sensitivity", N = 68, alpha = 0.05, tail
= 1, power = .80, theme = "pr")
```

The resulting graph is shown in Figure 4.

**Figure 4**. Example of a *sensitivity* power analysis graph with PLS-SEM-power (publication-ready graph)

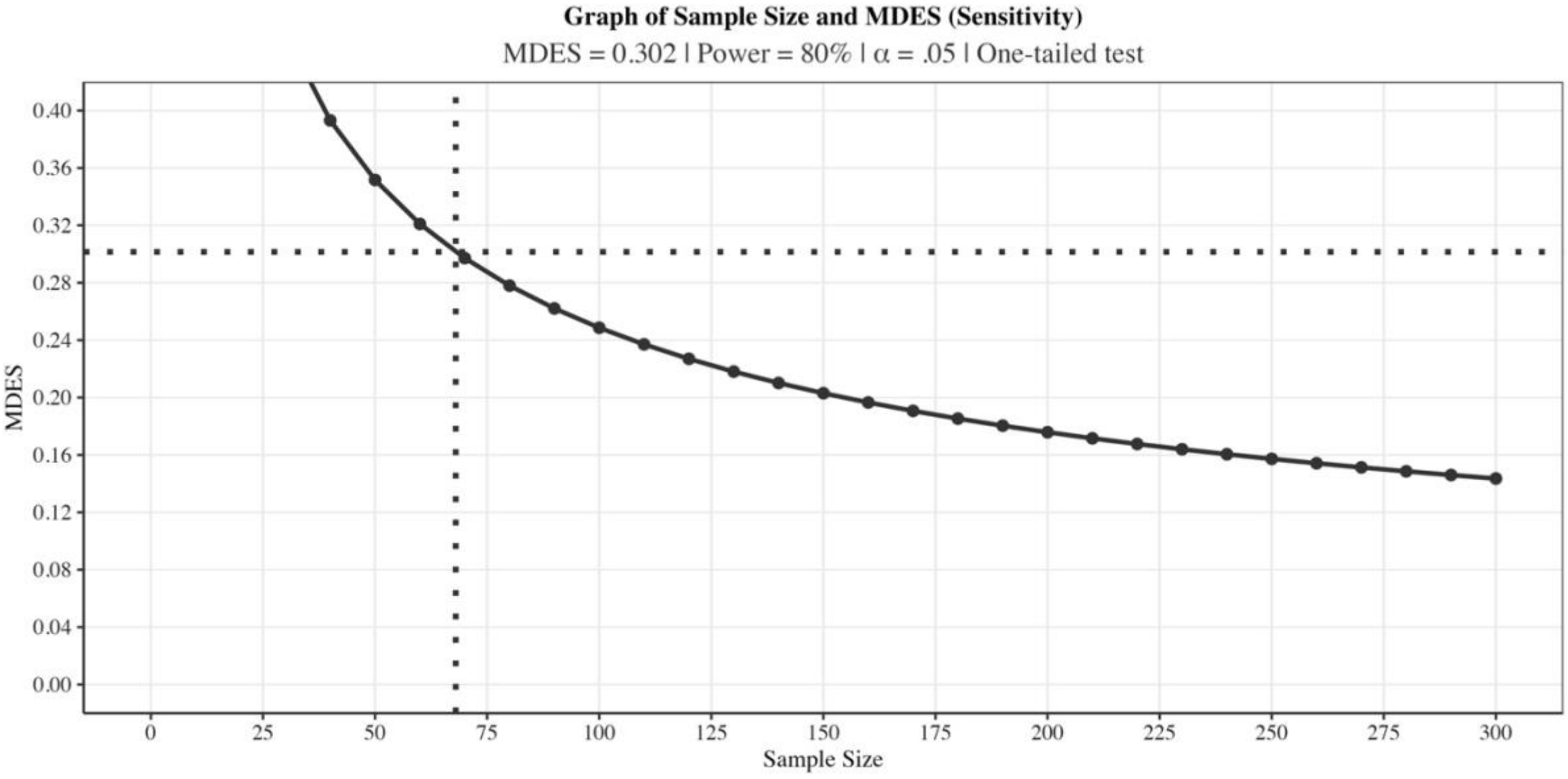

# Limitations and Special Considerations for Small Sample Sizes

Despite its simplicity, conservativeness, and practical advantages for sample determination in PLS-SEM, the inverse square root method has its negative sides, and some precautions must be taken for a more informed and solid usage. While it is straightforward and widely applicable, it assumes relatively simple model structures and does not account for complexities such as multi-layer constructs or formative indicators. Furthermore, with this method, the whole power analysis is based on one single path coefficient and not on the size of the most complex regression in the (formative) models or on the size of the overall model. This also implies, somewhat inevitably, that all the path coefficients below the value on which the power analysis is based will be, *de facto*, underpowered. For this reason, it has been suggested that the researcher should explore a range of MDES (Hair et al., 2021, p. 18).

It is also worth stressing that the choice between a one- or two-tailed test should be based on the substantive hypothesis being tested (Kock, 2015). *PLS-SEM-power* provides both options, but this flexibility does not alter the underlying assumptions and limitations of the inverse square root method.

Lastly, the *inverse square root method*, as discussed by Kock and Hadaya (2018), is recommended when the sample size ($N$) exceeds 10. For sample sizes smaller than or equal to 10 ($N \leq 10$), the *gamma-exponential method* is more appropriate because it accounts for the distribution of path coefficient estimates in small-sample scenarios. For values of N greater than 10, the *inverse square root method* and the *gamma exponential method* yield substantially equivalent results. However, when working with very small samples, it is advisable to use the gamma exponential method to obtain more accurate estimates.

# Conclusions

This Shiny App and R package simplify both *a priori* (sample size estimation given MDES) and *sensitivity* (MDES estimation given sample size) power analyses for PLS-SEM by interactively implementing the inverse square root method formula from Kock and Hadaya (2018) using one- or two-tailed calculations. The tool is designed to assist researchers in swiftly evaluating the feasibility of their designs and the detectability of particular effect sizes.

Thanks to R and Shiny's open-source nature, this application can be adapted or extended to suit different methodological needs. Hopefully, it will promote better planning and reporting of PLS-SEM analyses, reducing the risk of underpowered studies and providing more reliable findings.

# Declarations

## Funding

The author gratefully acknowledges the Research Council of Finland for supporting the Centre of Excellence in Music, Mind, Body and Brain (grant no. 346210) and the FRAMES project (grant no. 368151), which provided the broader academic and professional environment in which this software was independently developed. The software was not developed as a deliverable of either project.

## Conflicts of interest

The authors declared no potential conflicts of interest concerning the research, authorship, and/or publication of this article.

## Shiny App

The PLS-SEM-power Shiny App is freely available at the following link:
https://aleansani.shinyapps.io/pls-sem-power/

## Code availability

The R code and a user guide of the *plssempower* R package are freely available on GitHub at the following link: https://github.com/AleAnsani/plssempower.